\documentclass[article,aps,pra,twocolumn]{revtex4}  

\usepackage{amsmath}    
\usepackage{graphicx}   
\usepackage{float} 
\usepackage{tabulary} 

\newcolumntype{C}{>{\centering\arraybackslash}p{0.22\columnwidth}}
\newcolumntype{M}{>{\centering\arraybackslash}m{0.22\columnwidth}}
\newcommand{\ket}[1]{\left|#1\right\rangle}
\newcommand{\bra}[1]{\left\langle#1\right|}

\usepackage[binary]{SIunits}
\usepackage{xcolor}

\usepackage{chemformula} 

\begin{document}

\title{Detector blinding attacks on counterfactual quantum key distribution}
\author{Carlos Navas Merlo} 
\affiliation{Universidad de Valladolid}
\author{Juan Carlos Garcia-Escartin}
\email{juagar@tel.uva.es}
\affiliation{Universidad de Valladolid, Dpto. Teor\'ia de la Se\~{n}al e Ing. Telem\'atica, Paseo Bel\'en n$^o$ 15, 47011 Valladolid, Spain}
\date{\today}

\begin{abstract}
Counterfactual quantum key distribution protocols allow two sides to establish a common secret key using an insecure channel and authenticated public communication. As opposed to many other quantum key distribution protocols, part of the quantum state used to establish each bit never leaves the transmitting side, which hinders some attacks. We show how to adapt detector blinding attacks to this setting. In blinding attacks, gated avalanche photodiode detectors are disabled or forced to activate using bright light pulses. We present two attacks that use this ability to compromise the security of counterfactual quantum key distribution. The first is a general attack but technologically demanding (the attacker must be able to reduce the channel loss by half). The second attack could be deployed with easily accessible technology and works for implementations where single photon sources are approximated by attenuated coherent states. The attack is a combination of a photon number splitting attack and the first blinding attack which could be deployed with easily accessible technology. The proposed attacks show counterfactual quantum key distribution is vulnerable to detector blinding and that experimental implementations should include explicit countermeasures against it.
\end{abstract}
\maketitle

\section{Counterfactual Quantum Key Distribution}
Quantum information processing and quantum communication protocols offer new capabilities beyond what is possible with classical systems \cite{NC00}. In particular, Quantum Key Distribution (QKD) allows two parties, Alice and Bob, to use an insecure channel to agree on a secret private key \cite{SBC09}. 

In all QKD protocols, Alice and Bob make secret random choices and use quantum systems to agree on some common knowledge. By using quantum systems, usually a bit 0 or 1 mapped into some physical property like the polarization of a photon, they can discover any intruder. An eavesdropper, Eve, will change the system during the act of measurement. After a public discussion through an authenticated channel, available for everyone, Alice and Bob can compare their measurement statistics and estimate the information Eve might have gained. With this estimation, they can distill a smaller secret key leaving out any information Eve might have learnt about their chosen bits or, at least, find out their key is too compromised and cannot be used. This whole procedure that leaves out transmission errors and the influence of Eve is known as secret-key distillation \cite{Ass06}.

There is an interesting group of QKD protocols which are based on counterfactual schemes inspired by the Elitzur-Vaidman bomb test \cite{EV93} and interaction-free measurement \cite{KWH95}. The reference protocol in this family is Noh's counterfactual QKD protocol \cite{Noh09}. As opposed to other QKD proposals like BB84 \cite{BB84}, B92 \cite{Ben92}, SARG04 \cite{SARG04} or some versions of E91 \cite{Eke91}, there is a part of the quantum state used in the protocol that never leaves Alice's side when Alice and Bob agree on their choice. 

In this paper, we study detector blinding attacks in which Eve uses bright light pulses to control the detectors of Alice and Bob \cite{SLA11,LWW10a} and show that, while counterfactual QKD makes blinding attacks more difficult to carry out, they can be modified to be successful and should be taken into account. 

\subsection{Protocol under attack (N09)}
\begin{figure}[H]
	\centering
	\includegraphics[width=0.9\linewidth]{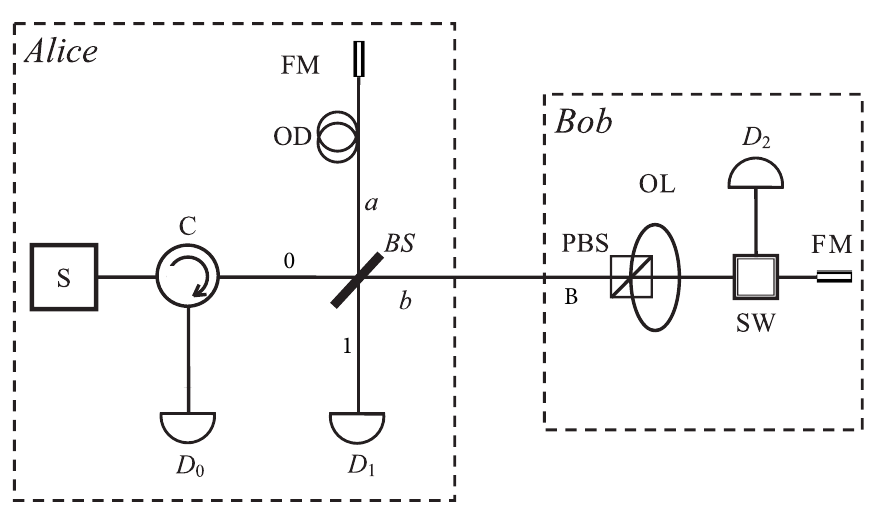}
	\caption{Implementation of the N09 QKD protocol. Two sides in closed labs, Alice and Bob, communicate through an insecure channel. A source $S$ sends a photon in a random polarization ($H$ or $V$) to a circulator $C$ which directs it to a beam splitter $BS$. One arm out of the beam splitter remains at Alice's lab (a) and the other goes into the public channel (b). In Bob's side, the photon goes through a polarizing beam splitter $P\!B\!S$ and an optical delay line $OL$ that allows an electrically-controlled switch $SW$ to choose which polarization is directed towards a detector $D_2$. Alice and Bob have Faraday mirrors $F\!M$ which reflect the photons back into the $BS$. The change in polarization is ignored for simplicity. Alice also has an optical delay $O\!D$ which guarantees the interferometer is properly balanced and both paths have the same length and losses. After the $BS$ the photon goes either into arm 0 where it reaches detector $D_0$ after going through $C$, or into arm 1, leading to detector $D_1$.}
	\label{fig:1}
\end{figure}

We consider the counterfactual QKD protocol N09 of Noh \cite{Noh09}. Figure \ref{fig:1} shows the basic setup. We assume the whole system is implemented over optical fiber, but it can be adapted to free space communication \cite{BCD12}. 

The whole system is a distributed Mach-Zehnder interferometer with two sides. A transmitter, Alice, has a source $S$ which produces a single photon either in horizontal or vertical polarization at predefined time windows. We consider the laser emits a diagonally polarized photon and there is an electrically controlled polarization rotator where the choice of the final polarization comes from a true random number generator. The photon goes to an optical circulator $C$ which sends it to a beam splitter $BS$ with a reflectivity $R$ and a transmissivity $T=1-R$.

The source prepares an input state $\ket{P}_0\ket{0}_1$ where $\ket{P}_0$ denotes a photon in the upper arm input either with horizontal or vertical polarization ($P=H$ or $P=V$ respectively) and $\ket{0}_1$ is the vacuum state of the lower arm. At the beam splitter we have the evolution
\begin{equation}
\ket{P}_0\ket{0}_1\to \sqrt{R}\ket{P}_a\ket{0}_b+i\sqrt{T}\ket{0}_a\ket{P}_b
\end{equation}
where the upper arm stays inside Alice's lab (path $a$) and the lower arm goes to the insecure channel that leads to Bob (path $b$).

At the other side of the channel, Bob chooses a random polarization, horizontal or vertical, and uses a polarizing beam splitter $P\!B\!S$ to separate the paths of the incoming photons in each polarization. He has an optical delay line, $OL$, that is combined with a switch $SW$ to direct the photons in the chosen polarization to a photon detector $D_2$. A photon in the orthogonal polarization is not measured. It will go to a Faraday mirror $F\!M$ and be reflected back to the channel. 

In order to keep interference, Alice has a similar setup in her side, with an optical delay, $O\!D$, which matches the delay and losses of the path including the communication channel and Bob's setup. Alice also has a Faraday mirror at the end of her fiber delay line. We consider reflection from Bob comes with a $\pi$ phase shift.

The Faraday mirrors rotate the polarization by $\pi/2$ on reflection. This is useful in practical fiber setups, where they compensate birefringence effects like thermally-induced polarization mode dispersion. A different polarization is used on the way back so that the channel effect on the photon is the same for both polarization choices. The effects of thermal drift and similar variations, which can change the channel behaviour for different polarizations over time, happen at much larger timescale than the roundtrip time and the compensation with Faraday mirrors is an effective correction. 

While there will be a change of polarization on reflection in practical implementations, we will ignore it in the analysis for simplicity. The required corrections are immediate. 

At Bob's side there are two possibilities. If Alice chooses a different polarization than the one Bob measures, the photon is reflected and, when it reaches Alice, there is a constructive interference for the output that leads to $D_1$ and a destructive interference for the output that reaches $D_0$.
The evolution of the shared quantum state, with the corresponding sign change upon reflection from Bob, is
\begin{eqnarray}
\sqrt{R}\ket{P}_a\ket{0}_b-i\sqrt{T}\ket{0}_a\ket{P}_b &\to& \nonumber\\
R\ket{P}_0\ket{0}_1+i\sqrt{RT}\ket{0}_0\ket{P}_1\nonumber\\
-i\sqrt{RT}\ket{0}_0\ket{P}_1 +T\ket{P}_0\ket{0}_1\nonumber\\
&=&\ket{P}_0\ket{0}_1.
\end{eqnarray}
This happens with a total probability of 1/2, the probability Alice and Bob random choices are different. Interference at the beam splitter guarantees the only detector that clicks is $D_0$.

If Bob measures in the same polarization Alice prepared her photon, the superposition in the state $\sqrt{R}\ket{P}_a\ket{0}_b-i\sqrt{T}\ket{0}_a\ket{P}_b$ is destroyed. With a probability $T$, Bob finds a photon in $D_2$ (with a total probability of $\frac{T}{2}$, including the 1/2 probability of choosing the same polarization). The photon is lost and Alice gets no clicks at her detectors. If Bob finds no photon in $D_2$, with probability $R$, the photon is with certainty inside Alice's lab. Now, the beam splitter has a vacuum state in the channel arm and the evolution is
\begin{equation}
\ket{P}_a\ket{0}_b\to \sqrt{R}\ket{P}_0\ket{0}_1+i\sqrt{T}\ket{0}_0\ket{P}_1.
\end{equation}
The total probabilities for detections inside Alice are $\frac{R^2}{2}$ for $D_0$ and $\frac{RT}{2}$ for $D_1$, including the $R/2$ probability that Bob measures in the same polarization Alice has sent her photon, but gets no photons in his measurement.

In the QKD protocol, Alice and Bob exchange $N$ photons and record their configuration and their measurement results. In order to detect any eavesdropper and to establish a common secret key, for each photon they reveal which detectors, if any, clicked. When $D_0$ or $D_2$ have fired, Alice and Bob also reveal their chosen polarization to check for eavesdroppers. If the statistics are different from expected, they can discover there has been some tampering in the channel. When only $D_1$ clicked both Alice and Bob know their own secret polarization choices and that they must be orthogonal. For these results they can agree on a bit value, for instance 0 when Alice chose horizontal polarization and 1 when she chose vertical. They also can use the frequency of clicks in $D_1$ to check for an eavesdropper and make public a few polarization choices to check they, in fact, chose the same values. Any action of the eavesdropper will be discovered at this stage.

The name counterfactual comes from how the bits of the key are generated. Alice and Bob only agree on the bit value when Bob could have found a photon in $D_2$ but didn't. The protocol only works if Bob has the possibility to find a photon in $D_2$ so that the measurement produces a change in the shared state. However, all the bits of the key are generated when, after the measurement, we know for sure the photon never left Alice's lab. In that respect, the protocol seems to have an added security factor. Eve cannot measure the channel without either destroying the superposition and forcing a vacuum state inside Alice's lab or finding a vacuum state in the channel. In the first case, Eve cannot modify Alice's part of the state and, in the second, Eve does not know Alice's chosen polarization. In both cases Eve cannot produce new quantum states which will produce the same detector statistics as the eavesdropper-free channel.

\subsubsection{Security proofs and quantum hacking}
For the original assumptions, the N09 protocol has a security proof showing it is equivalent to an entanglement distillation protocol \cite{YLC10}. However, this proof only works for a perfect, lossless channel. For lossy channels, there is an attack which compromises any system where the link between Alice and Bob has more than $3~\deci\bel$ losses \cite{LWL14,Li14}. There is also a proof of the security of the protocol against general collective attacks when the source generates attenuated coherent states instead of single photons \cite{YLY12}, which includes channel losses. In this setting there is a reduction in the amount of private information in each exchange between Alice and Bob, but they can still generate a common secret bit sequence.

We will present an attack valid for both settings, with single photons and weak coherent states, and show it only needs moderate resources against counterfactual QKD with weak coherent states. 

Attacks to quantum key distribution can be classified in two large families. On one hand there are explicit strategies to attack within the protocol assumptions, like in the intersend-and-resend attacks \cite{BBB92} or photon number splitting attacks \cite{BLM00}. Security proofs establish the limits of these attacks and give bounds to the information Eve can learn \cite{LCT14}. With these guaranteed bounds, Alice and Bob can use key distillation methods that generate a smaller private key and the protocol is considered secure \cite{Ass06}.

In practice, any physical implementation will present side channels and imperfections which can be leveraged for an attack \cite{HBA18}. The most successful attacks to QKD use physical properties of the system which are not properly modelled in the protocol assumptions. Quantum hacking takes advantage of these deviations to learn additional information or otherwise alter the experimental setups inside Alice or Bob's labs \cite{LWW10a, LWW10b,LSM11, LAM11, WLW11,QFL07,CSL17,ZFQ08,SRK15,FQT07,XQL10,SCB15}. 

For counterfactual QKD, a prominent line of attack is using undetected light to peek inside Bob's lab and learn his configuration. If Eve can discover Bob's chosen polarization without changing the global detection statistics, she can break the security of the whole protocol. Alice and Bob do not know the other side's choice of polarization and the secret key is established from the knowledge of the locally chosen bit and the public communication stage. Eve can always listen to this communication and, if she knows the polarization choice of either side, she can produce the bits from the key exactly as well as that side can.

One way to probe inside Bob's lab is using a counterfactual Trojan horse \cite{WZT12}. Eve can reproduce Alice's setup and send a polarized photon to Bob with the same properties of those from Alice (so they cannot be filtered). However, in order not to be detected, Eve needs to use techniques inspired from interaction-free measurement \cite{KWH95,KWM99} and generate superpositions with a small enough probability of producing an actual detection. This requires the input from Bob to be open and measuring for a longer time than strictly necessary to measure Alice's photons.

A second proposed attack is using ``invisible'' photons which can be absorbed during measurement but will never produce a click at the detector \cite{YWM16}. If Eve monitors the channel she can tell if the photon was absorbed or not and find out the measurement configuration inside Bob. One possible way to have these invisible photons is choosing a small delay and use polarized photons at the same wavelength as Alice. Photon detectors tend to work in a gated mode and if the photon arrives before or after the detector's active window, it will not trigger a detection. This is possible when the switch inside Bob does not change its configuration for a time longer than the time bin of the active detector. Another option is using light at a wavelength outside the detection range from Bob's detector. The light could even be classical. By checking the delay in the outcoming light Eve can deduce Bob's measurement choice. Bob can prevent this attack by installing wavelength filters which only let in photons in a narrow frequency band.

In this paper we look at {\bf detector blinding} attacks \cite{SLA11,LWW10a}. Many quantum communication systems with single photons or weak coherent states use avalanche photodiodes APDs as detectors. Single photon detection happens in what is called the Geiger mode, where the diode works under a strong reverse bias (above a breakdown voltage). In this mode of operation, thermally generated carriers can trigger an avalanche and produce false detections which are known as \emph{dark counts}. After a detection there is also a dead time in which the APD needs to recover and cannot detect any photon. 

For these reasons, the preferred operation is usually a gated regime where most of the time the APD is in the linear mode from which it can easily change to the Geiger mode for the short periods of time when an incoming photon is expected. In the linear mode the output current is proportional to the power of the incoming light. Detector blinding attacks use bright light to switch from the Geiger to the linear mode, where single photon events do not register. Eve can either suppress all the counts in a detector or, if she uses an additional pulse trigger pulse over a threshold, force a detection at will. 

This attack is quite general and has been demonstrated for many of the usual gated configurations used in a variety of QKD schemes  \cite{LJW11,GLL11,LLK14,QKM18,CHE19,SSL20}.

\section{Model of the channel and protocol assumptions}
In the main attack we will consider a realistic implementation under the following assumptions:
\begin{itemize}
\item[-]{\emph{Source with weak coherent states:}}
Instead of single photons, the source produces coherent states which are a superposition of number states $\ket{n}$ with $n$ photons such that:
\begin{equation}
\label{coherent}
\ket{\alpha}=e^{-\frac{|\alpha|^2}{2}}\sum_{n=0}^{\infty}\frac{\alpha^n}{\sqrt{n!}} \ket{n}.
\end{equation}
In coherent states, the measured photon number follows a Poisson distribution with a mean photon number $|\alpha|^2$ for a coherent state $\ket{\alpha}$. The mean photon number is chosen to be smaller than 1. A typical value $|\alpha|^2=0.1$ is often used to minimize the probability of having multiple photons in the channel.

This is a usual approach for QKD in general. Attenuated laser light produces these weak coherent states at a fraction of the cost of single photon sources with a comparable generation rate \cite{EFM11} and the single photon condition is usually relaxed. 

For the limited number of experimental demonstrations of counterfactual QKD, while it is possible to use heralded single photon sources to realize the original protocol \cite{BCD12}, the bit generation rates are more limited than implementations using attenuated coherent states \cite{RWW11,LJL12}. However, departing from the original protocol means a reduction of security which must be accounted when extracting the key to keep within the constraints of the security proof for weak coherent states \cite{YLY12}.

\item[-]{\emph{Binary imperfect detectors:}}
We assume the detectors have a limited efficiency $\eta<1$ and have no photon-number-resolving capability. This is true of the APD detectors we consider in our attack. For instance, in a typical optical fiber implementation, the photons usually have a wavelength of 1550 $\nano\meter$ (the band at which standard fiber attenuation is the lowest). Infrared APD detectors are notoriously inefficient and typical efficiencies are around $\eta=0.1$, which can be increased to about $\eta=0.25$ at the price of increasing the dark count rate \cite{EFM11}. 

We also consider all-or-nothing detection. An APD shows an output current whenever there is an avalanche, which is equally triggered by one or multiple photons. This is a good approximation to a measurement in the $\left\{\ket{0}\bra{0}, I-\ket{0}\bra{0}\right\}$ basis, where $\ket{0}$ is the vacuum state. We suppose the effiency is the same for both polarizations and will ignore the effect of dark counts (which could be added to the model if necessary). APDs have some limited photon-number-resolving capabilities, but the processing is usually complex and the results show limited accuracy. 

\item[-]{\emph{Lossy channels:}}
The photons can get lost in the channel between Alice and Bob. We consider channels with a transmission $\sigma<1$. In standard optical fiber, a typical loss value at $1550~\nano\meter$ is $0.2~\deci\bel / \kilo\meter$.

\item[-]{\emph{The detectors can be partially controlled by an attacker:}}
We consider a blinding attack with bright light where we can either send bright light to blind a detector (reducing any clicks to 0) or, by injecting a trigger pulse on top of the bright light, make it click with certainty. This will always work for Bob, but for Alice's detectors, depending on the reflection and transmission parameters of her beam splitter, we might have more limited options. From Figure \ref{fig:1} we can see that we can (i) blind both detectors, (ii) make both detectors click, or (iii) either blind $D_0$ and make $D_1$ click, if $T<R$, or blind $D_1$ and make $D_0$ click if $T>R$ for the right threshold in the trigger pulses. For $T=R=0.5$ we have the same effect on both detectors.

\end{itemize}

This paper considers how the blinding attacks demonstrated for other systems translate to counterfactual QKD. While such an attack is possible, it is not direct as we need to deal explicitly with the photons that remain inside Alice. This is the new element considered in our attack. All the previous assumptions for realistic implementations of counterfactual QKD are already covered in a security proof \cite{YLY12} and can be accounted for. 

\section{Blind and reduce losses attack}
The first attack we present is general and will work both for the single photon and the weak coherent state versions of the protocol. The basic principle is using blinding light in one polarization to learn Bob's polarization choice. 

If Eve produces a strong blinding signal in one polarization chosen at random and monitors the signal coming back from Bob, she knows whether her choice is the same as Bob's (the light doesn't come back and is directed towards Bob's detector, which is blinded) or different (the light is reflected and goes back to the channel).  

\begin{figure}[t]
	\centering
	\includegraphics[width=0.9\linewidth]{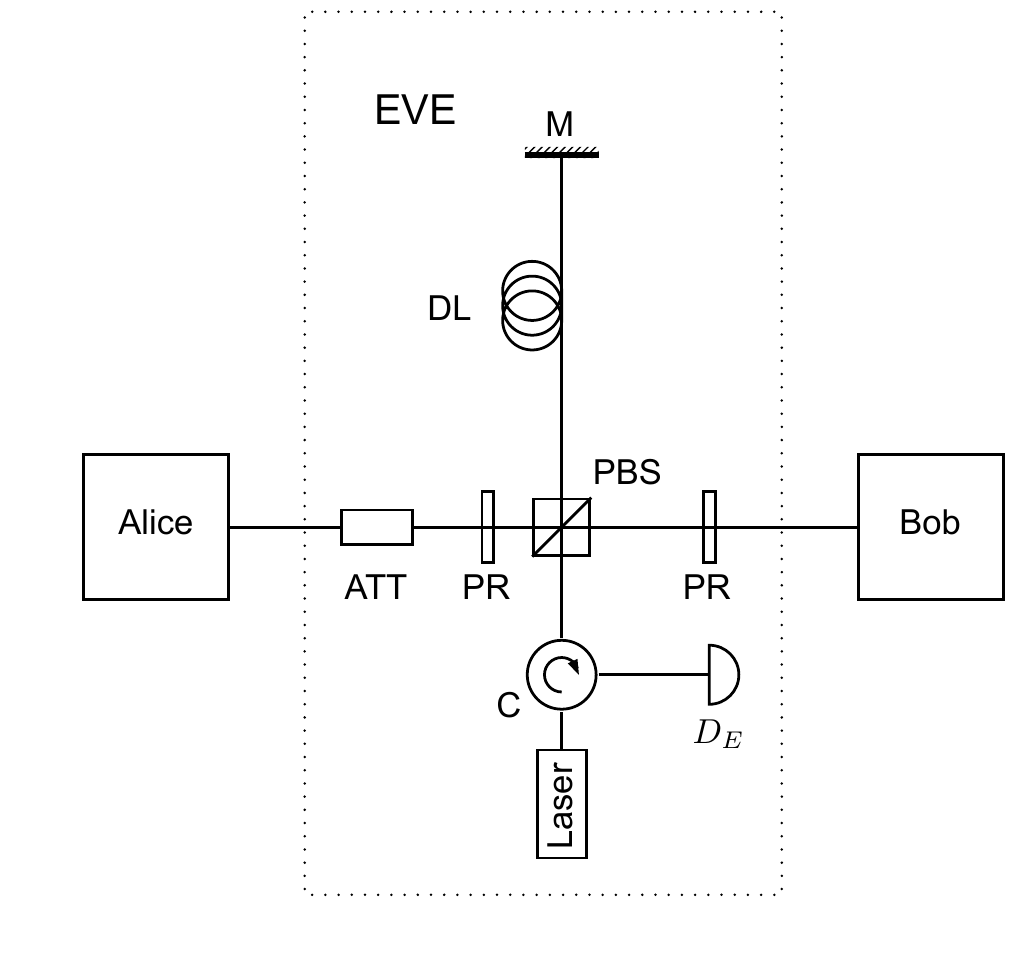}
	\caption{Setup for the blinding attack. Eve replaces the channel for an interferometer which combines her bright blinding light in one polarization with the orthogonal polarization coming from Alice using a polarizing beam splitter, PBS, and series of electronically controlled polarization rotators, PR, which allow her to change her choice. The original signal from Alice at Eve's chosen polarization is kept stored in the arm of the interferometer that is not going to Bob with a delay line $DL$ and a mirror $M$. After going inside Bob's lab, the PBS restores Alice's pulse in the delay line $DL$ into the channel and extracts the blinding pulse which goes to a circulator $C$ which directs it to Eve's detector $D_E$. Eve can learn the polarization of Bob's measurement by checking if she gets any returning light (different polarization as Eve's pulse) or not (same polarization). With probability one half, the bright light pulse does not return and Bob is blinded. In that case, Eve sets her attenuator $ATT$ to high extinction to suppress any photons from Alice. When the bright pulse returns, Eve has learnt Bob's measurement choice and the protocol continues as expected. In order to keep the detection statistics at Bob, Eve must reduce the losses in her channel to Bob and the delay arm to one half of their expected value. In order to keep the detection statistics at Alice, she must later adjust her attenuator $ATT$ to make up for the reduced loss and restore the expected total loss for the roundtrip channel.}
	\label{fig:2}
\end{figure}

Eve needs to guarantee the count statistics are not disturbed. A simple setup to achieve that is shown in Figure \ref{fig:2}. Eve can take Alice's photons out of the channel and into a polarizing beam splitter, PBS, which reflects vertical polarization and transmits horizontally polarized light. On the other input of the PBS Eve injects a blinding signal in vertical polarization. Alice's vertically polarized photons are kept in a delay line. 

If Bob is measuring vertically polarized light, his detector is blinded by Eve's bright light. If he is measuring horizontally polarized light, Eve's light will be reflected. Back at the PBS, Eve can measure for reflected light at $D_E$ and learn Bob's choice by checking whether any light comes out (horizontal measurement) of not (vertical). If Bob and Eve chose a different basis the protocol follows as expected. Alice's photons either go to Bob and are subject to measurement (firing $D2$ or not with the expected statistisc) or they are kept in the delay line and go back to Alice to produce the expected interference. If Bob and Eve choose the same basis, the photons on the delay line must be blocked (they should have been measured by Bob). Eve has an electronically controlled attenuator (ATT in the Figure) which she can use to effectively destroy these photons by applying a large attenuation. This will perturb the expected statistics, but it can be masked in the channel's loss. 

The whole setup for Eve includes two polarization rotators PR that allow her to repeat the procedure injecting bright horizontal light pulses and passing the vertical light from Alice to Bob. For electrically activated PRs, like Pockels cells, Eve chooses which polarization goes throught and which is kept in the delay line. In both cases she can learn Bob's configuration.

Eve can hide her presence with two measures. First, she chooses at random the polarization she injects. The net effect is that she must block half the photons in the channel. When Eve and Bob choose the same polarization she must make sure no detector clicks. She doesn't know Alice's polarization and cannot reproduce her statistics. The resulting $3~\deci\bel$ increase in the channel loss presents a problem. Even if the excess loss could be attributed to natural channel variation, in counterfactual QKD, Alice keeps inside her lab part of the state and needs to have a tight estimation of the channel loss, which she reproduces inside her lab. An additional one half loss is too large for counterfactual QKD.

In order to reproduce the statistics at Alice and Bob, Eve must perform a second correction: she can replace the channel from Alice to Bob with a channel with half the losses so that Bob's detection statistics remain as expected. On the way back, for the photons that reach Alice, Eve must use her attenuator to compensate for the smaller loss on the way to Bob. Blocking half the photons already introduces a $3~\deci\bel$ loss, but Eve must add an extra $3~\deci\bel$ to account for the $6~\deci\bel$ loss reduction in the roundtrip channel, with half the losses in each direction. The result is that the state going back to Alice has suffered the same total loss as if Eve had not hijacked the channel. 

In principle, this is possible. For instance, Eve can replace the fiber channel from Alice to Bob with a fiber with almost no loss or with a vacuum free space channel. However, in practice, this would be too technologically demanding. 

In order to be able to exploit blinding with current technology, we propose a more realistic attack which combines the blinding strategy with a photon number splitting attack valid for counterfactual QKD systems using weak coherent states. With the additional information from direct measurement of Alice's state, Eve can mount succesful blinding attacks where she only needs to reduce the channel loss by a small amount (if it is needed at all).

We assume Eve is restricted to current ultra low loss optical fibers, which can achieve losses below $1.5~\deci\bel/\kilo\meter$ at telecom wavelengths around $1550~\nano\meter$ \cite{Ten16,TSM17,Tam18,TSM18}. The difference between Eve's channel and the usual channel with typical single mode fiber, with a 0.2 $\deci\bel/\kilo\meter$ loss at 1550 $\nano\meter$, is enough for an effective attack.

\section{Combined blinding, measurement and faked states attack}
The blinding attack from the previous Section can be refined if Alice and Bob are using coherent states. In that case, the global state used to establish the key is no longer entangled. For a polarization $P$ (where $P$ can be horizontal, $H$, or vertical, $V$, polarization), the state is no longer $\sqrt{R}\ket{P}_a\ket{0}_b+i\sqrt{T}\ket{0}_a\ket{P}_b$ but $\ket{\sqrt{R}\alpha,P}_a\ket{i\sqrt{T}\alpha,P}_b$ for coherent states $\ket{\alpha,P}$ following Eq. (\ref{coherent}) with all the photons in the same polarization $P$. 

Now, a measurement in the channel will not affect the first half of the state, which remains $\ket{\sqrt{R}\alpha,P}_a$. If Eve measures one or more photons in the coherent state $\ket{i\sqrt{T}\alpha,P}_b$ that goes to the channel, she can determine the polarization chosen by Alice. For this bit she has complete information. She can just put into the channel a new $\ket{i\sqrt{T}\alpha,P}_b$ state and let the protocol run normally with her side information. In our protocols, depending on her side information, Eve will introduce different faked states in the channel \cite{MH05} so that the final detector statistics are indistinguishable from those of an eavesdropper-free channel.

However, the amplitude is chosen so that the mean photon number in the channel $T|\alpha|^2<1$. Finding a photon in the channel should be unlikely. If Eve measures the vacuum state, she doesn't know which state she should send to Bob. We will show a strategy to overcome this problem by blinding Bob's and Alice's detectors. 

First we assume Alice and Bob do not have specialized detectors, i.e. they just measure the presence or absence of photons, but do not distinguish between horizontal and vertical polarization. Later we will adapt the attack for polarization-discriminating detectors (Alice, Bob or both have a polarizing beam splitter and two detectors for each measurement so that they can tell apart horizontally and vertically polarized photons).

\subsection{Expected count statistics}
For a successful attack we ask two conditions. First, Eve must know Alice or Bob's choice of polarization (or both). Second, the count statistics at $D_0$ and $D:1$ (Alice's lab) and $D_2$ (Bob's lab) are not modified due to Eve's actions. 

Eve can figure out the expected statistics from the unperturbed channel's transmission $0<\sigma<1$ and her knowledge of Alice's configuration (the choice of the coherent state amplitude $\alpha$ and the tranmission of her beam splitter $T$). She also knows the efficiencies $\eta_0$, $\eta_1$ and $\eta_2$ of the detectors $D_0$, $D_1$ and $D_2$, respectively. All these details are part of the protocol and can be considered public under Kerckhoffs' principle (the security must not depend on keeping part of the protocol or the device secret, only on a secret value) \cite{Ker83}.

The initial state from Alice is
\begin{equation}\label{eq:4}
\begin{aligned}
\ket{\alpha,P}&= & e^{-\frac{1}{2}|\alpha |^2}\sum_{n=0}^{\infty}\frac{\alpha^{n}}{\sqrt{n!}}\ket{n,P}
\end{aligned}
\end{equation}
for a mean photon number $\langle n\rangle = |\alpha|^2$. At the beam splitter we have the evolution \cite{Lou00}:
\begin{equation}\label{eq:5}
\begin{aligned}
\ket{\alpha,P}_{0}\ket{0}_{1}& \to & \ket{\sqrt{R} \alpha,P}_{a}\ket{i\sqrt{T} \alpha,P}_{b}
\end{aligned}
\end{equation}
with $R+T=1$.

After crossing the channel from Alice to Bob, the coherent state that reaches Bob is $\ket{i\sqrt{T}\sqrt{\sigma}\alpha,P}_{b}$. 

If Bob measures in the same polarization Alice prepared her state, he will find a photon with probability 
\begin{equation}
\label{PD2}
{P_{D2}}(n>0)=\frac{1}{2}(1-e^{-\eta_2\sigma T |\alpha|^2})
\end{equation}
which includes the 1/2 probability of Alice and Bob making the same polarization choice. Regardless of the measurement result, the second crossing at Alice beam splitter happens for an input state $\ket{\sqrt{R} \sigma \alpha,P}_a\ket{0}_b$ which gives an output $\ket{R \sigma \alpha,P}_0\ket{i\sqrt{RT}\sigma \alpha,P}_1$ and count statistics
\begin{equation}
\label{PD0}
{P_{D0}}(n>0)=\frac{1}{2}(1-e^{-\eta_0\sigma^2 R^2  |\alpha|^2})
\end{equation}
and
\begin{equation}
\label{PD1}
{P_{D1}}(n>0)=\frac{1}{2}(1-e^{-\eta_1\sigma^2 R T |\alpha|^2}),
\end{equation}
again including the 1/2 probability of having the same polarization. 

The only case where a key bit is generated is when $D_1$ clicks. $P_{D1}$ gives the expected key bit rate (before key distillation). In the other scenarios, Alice and Bob announce their chosen polarization to check the statistics are correct. During key sifting, they might also disclose a fraction of the polarization choices when $D_1$ clicks in order to estimate the total qubit error rate and to check for the presence of an eavesdropper.

If Bob measures in the opposite polarization from Alice's state, the coherent state suffers a $\pi$ phase shift and travels back to her. At the beam splitter we have the interference of two coherent states for an input
\begin{equation}
\ket{\sqrt{R}\sigma \alpha,P}_{a}\ket{-i\sqrt{T}\sigma \alpha,P}_{b}
\end{equation}
which becomes
\begin{equation}
\ket{\sigma  \alpha,P}_{0}\ket{0}_{1}.
\end{equation} 
In this case, $D_1$ never fires and the resulting coherent state goes to $D_0$ with certainty producing a click with a probability:
\begin{equation}
\label{PpD0}
{P'_{D0}}(n>0)=\frac{1}{2}(1-e^{-\eta_0\sigma^2 |\alpha|^2})
\end{equation}
including the 1/2 probability for Alice and Bob choosing different polarizations. 

\subsection{Attack protocol without polarization detection}
\label{AttProt}
For the general attack, we assume Eve can measure the coherent state at Alice's output and learn its polarization if she finds at least one photon. We also suppose she can moderately improve the channel transmission (for a reasonable loss reduction below the 20\%), blind or force a click in Bob's detector at will and blind \emph{both} detectors in Alice's lab, which is true for any $T$ in Alice's beam splitter. 

In this section we outline Eve's decision tree for the general attack protocol and will discuss the expected detector probabilities and the necessary parameter optimization in the following sections.

First, Eve must decide whether she measures Alice's state or not with a fixed probability $x$ of measuring. There are two scenarios:

\subsubsection*{\textbf{Eve does not measure the $\ket{i\sqrt{T}\alpha}_b$ state}}

\textbf{Eve uses the blind and reduce losses strategy.} Eve follows the attack in the previous section but with a minimal intervention in the channel. We take as a reference a modified channel transmission $\sigma\leq \sigma'\leq 1.2 \sigma$. When Eve has chosen the same polarization Eve measures, instead of blocking Alice's pulse, she measures it in the same polarization Bob did, letting the orthogonal polarization pass. Eve will mimic the behaviour inside Bob and can, with a small probability, learn Alice's polarization. In both cases, the state Alice receives is exactly the same as in the eavesdropper-free protocol.

\subsubsection*{\textbf{Eve measures the $\ket{i\sqrt{T}\alpha}_b$ state from Alice}}
Here the strategy depends on the measurement's outcome:

\vspace{1ex}
\textbf{Eve's detectors finds one or more photons.} Now Eve knows Alice's polarization choice. Eve sends Bob a bright light pulse in the same polarization Alice chose in order to force a detection with probability $y$ when Bob measures in Alice's polarization or a blinding pulse with probability $1-y$. By monitoring the light back from Bob, Eve learns Bob's polarization choice. This is a case of \emph{perfect information}. Eve knows the polarization choices of both Alice and Bob.  

\begin{itemize}
\item If Bob measures in the same polarization as Alice's photons, Eve manages to force a click in $D_2$ for a fraction $y$ of all the pulses. Eve blinds Alice's detectors in order to compensate for the extra counts of the faked states in the rest of the protocol.

\item If Bob measures in the opposite polarization, Eve produces a faked state with the expected amplitude and polarization so that, back in Alice's beam splitter, there is a constructive interference and the right coherent state reaches $D_0$.
\end{itemize}

\textbf{Eve's detectors find no photons.} This is the worst case. Eve cannot learn Alice's polarization choice and she has destroyed the original state. In order to mask her presence, Eve:

\begin{itemize}
\item Sends diagonally polarized light to Bob to blind his detector (for both polarization choices) and measures the polarization that survives back to the channel to learn Bob's polarization choice.
		
\item With a probability $z$ Eve prepares a faked state for Alice with the expected amplitude but in the polarization orthogonal to Bob's choice. In that case:

\begin{itemize}
\item When Bob measures in the same polarization Alice has chosen, the photons inside Alice and the photons from Eve are orthogonal and there is no interference. The probability of detection in $D_0$ and $D_1$ will increase.
		
\item When Bob does not measure in the same polarization Alice has chosen, Eve's state is the expected state for the protocol and Alice obtains the expected constructive interference and the expected coherent state at $D_0$.
\end{itemize}

\item With a probability $1-z$ Eve blinds all the detectors inside Alice. Both $D_0$ and $D_1$ produce no counts.
\end{itemize}

Notice that, in all of the cases, Eve learns either Bob's polarization choice or both Alice's and Bob's choice. The only requirement for a successful attack is that the detector statistics cannot be told apart from the expected ones.

\subsection{\textbf{Expected detector statistics for weak coherent state counterfactual QKD under the combined blinding, measurement and faked states attack}} 

Eve's strategy can be adjusted by tuning the parameters $x\in[0,1]$, the fraction of light pulses from Alice that Eve measures, and $z\in [0,1]$, the probability Eve prepares a faked state when she measures the state from Alice and finds no photons. We assume Eve has a binary detector (with no photon resolving capabilities) with an efficiency $\eta_E$.

The detection rate in Bob's detector is
\begin{equation}\label{eq:17}
\begin{aligned} 
{{P}^{1}_{D2}}(n>0) &=\frac{1}{4}(1-x)(1-e^{-{\eta_2}{\sigma'}T|\alpha|^2})\\
&\quad +\frac{1}{2} x(1-e^{-{\eta_E}{T|\alpha|}^2})y.
\end{aligned}
\end{equation}
The main strategy is a blind and reduce losses attack (first term) combined with measurement (second term for a succesful measurement finding one or more photons). Eve blinds Bob's detector if she tries to measure the state from Alice and finds no photons.

Eve needs to compensate for the count reduction in $D_2$ due to blinding. In the blind and reduce losses attack Eve acts on the channel to Bob so that it has a transmission $\sigma'>\sigma$. Additionally, in the perfect knowledge scenario, Eve knows the polarization Alice chose. Eve can tune the probability $y$ of inducing a detection or sending a blinding pulse (with probability $1-y$). After checking the detection rate for a given $x$, Eve can make $y=1$ if the difference between the expected detection rate and the rate under attack is below $\frac{1}{2} x(1-e^{-{\eta_E}{T|\alpha|}^2})$ or choose the $y$ that makes the expected and the actual probabilities equal when that is possible.

The detection probabilities in Alice are studied for two scenarios:

\vspace{1ex}
- Alice prepares her coherent state in the {\bf same polarization} Bob measures:\vspace{1ex}

In this case, we must take into account that, with probability $x$, Eve measures the coherent state out of Alice. If she detects $0$ photons, she produces with a probability $z$ a faked state in a polarization orthogonal to Bob's choice with an amplitude $\sigma \sqrt{T} \alpha_\perp$. This state does not interfere with Alice's coherent state and they evolve independently at the beam splitter.

For the protocol we have given, the detection probabilities are
\begin{equation}\label{eq:15}
\begin{aligned} 
{{P}^{1}_{D0}}&(n>0) = \frac{1}{2}(1-x)(1-e^{-{\eta_0}{\sigma^2}R^2|\alpha|^2})\\
&\quad + \frac{1}{2}x e^{-{\eta_E}{T|\alpha|}^2}z(1-e^{-{\eta_0}{\sigma^2}(R^2|\alpha |^2+T^2| \alpha_\perp|^2)})
\end{aligned}
\end{equation} 
and
\begin{equation}
\begin{aligned} \label{eq:14}
{{P}^{1}_{D1}}&(n>0) =\frac{1}{2}(1-x)(1-e^{-{\eta_1}{\sigma^2}RT|\alpha|^2})\\
&\quad +  \frac{1}{2}xe^{-{\eta_E}T|\alpha|^2}z(1-e^{-{\eta_1}{\sigma^2}(RT|\alpha|^2+RT|\alpha_\perp|^2)}).
\end{aligned}
\end{equation}
The first terms assume a blind and reduce losses attack. The second terms give the counts when Eve uses faked states. The amplitude of these states could, in principle, be adjusted to give a further degree of freedom to Eve. In practice, we use the heuristic that the faked state should have the same amplitude as the expected state, which gives no errors when Alice and Bob polarizations are different ($\alpha_\perp=\alpha$). The faked state is only added if Eve measures the coherent state coming from Alice and finds zero photons. 

For equal amplitudes, the detection probabilities can be written as:
\begin{equation}
\begin{aligned} 
{{P}^{1}_{D0}}&(n>0) = \frac{1}{2}(1-x)(1-e^{-{\eta_0}{\sigma^2}R^2|\alpha|^2})\\
&\quad + \frac{1}{2}x e^{-{\eta_E}{T|\alpha|}^2}z(1-e^{-{\eta_0}{\sigma^2}(R^2+T^2)|\alpha |^2})\\
\end{aligned}
\end{equation} 
and
\begin{equation}
\begin{aligned} 
{{P}^{1}_{D1}}&(n>0) =\frac{1}{2}(1-x)(1-e^{-{\eta_1}{\sigma^2}RT|\alpha|^2})\\
&\quad +  \frac{1}{2}xe^{-{\eta_E}T|\alpha|^2}z(1-e^{-{\eta_1}{\sigma^2}2RT|\alpha|^2}).
\end{aligned}
\end{equation}

\vspace{1ex}
- Alice prepares her coherent state in a {\bf different polarization} from Bob's measurement:\vspace{1ex}

In all the strategies there are no clicks in $D_1$, which always gets the expected vacuum state (resulting from a destructive interference at the beam splitter) or is blinded. Similarly, except for the blinding pulses, $D_0$ is presented the expected state, either from a faked state or the originial state from Alice after going to Bob or Eve. The probability of detection is
\begin{equation}\label{eq:16}
\begin{aligned} 
{{P'}^{1}_{D0}}(n>0) &= \frac{1}{2}(1-xe^{-{\eta_E}{T|\alpha|}^2}(1-z))(1-e^{-{\eta_0}{\sigma^2}{|\alpha |}^2}).
\end{aligned}
\end{equation}

\subsection{Attack protocol against detectors with polarization discrimination}
\label{poldisc}
If Alice, Bob, or both can tell apart the polarization of their incoming photons, Eve needs a new strategy. In the original counterfactual QKD proposal polarization discrimination was assumed \cite{Noh09} and it is recommended to check for attacks. It can be implemented by replacing each detector with a polarizing beam splitter with a detector at both polarization outputs. We assume these detectors are also APDs without photon number resolving capabilities which can be blinded just as in the previous analysis.

We follow the same attack strategy of Section \ref{AttProt} but changing the behaviour when Eve measures Alice's state from the channel. Whenever Eve measures 0 photons in the coherent state from Alice, she must blind Alice's detectors in the polarization that is orthogonal to Bob's measurement choice. Otherwise, Alice could detect photons in $D_1$ when there should have been a destructive interference, which would expose Eve. As a result, in this attack version, $D_0$ has no counts in the cases where there would have been a constructive interference (Bob does not measure in Alice's polarization).  

When Bob measures in the same polarization Alice has chosen, the detection probabilities are exactly the expected: 
\begin{equation}\label{eq:19}
\begin{aligned} 
{P^{2}_{D0}}&(n>0) = \frac{1}{2}(1-e^{-{\eta_0}{\sigma^2}R^2|\alpha |^2})
\end{aligned}
\end{equation}
and
\begin{equation}\label{eq:18}
\begin{aligned} 
{P^{2}_{D1}}&(n>0) = \frac{1}{2}(1-e^{-{\eta_1}{\sigma^2}{RT|\alpha |}^2}).
\end{aligned}
\end{equation}
Here we have the expected statistics not only for the blind and reduce losses attack, but also for both scenarios in the measurement attack. For the perfect knowledge scenario, instead of blinding Alice, Eve does nothing reproducing the expected vacuum state. Similarly, when Eve measures Alice's state and obtains zero photons, the blinding light and the photons inside Alice are in orthogonal polarizations and everything runs according to the expected protocol (Alice only sees the state in her delay line). 

When Bob measures in a different polarization from Alice's choice, only $D_0$ can click. The detection rate is given by
\begin{equation}\label{eq:20}
\begin{aligned} 
{{P'}^{2}_{D0}}(n>0) &= \frac{1}{2} (1-xe^{-{\eta_E}{T|\alpha|}^2}) (1-e^{-{\eta_0}{\sigma^2}{| \alpha |}^2}),
\end{aligned}
\end{equation}
where $D_0$ is blinded when Eve tries to measure Alice's state from the channel but finds no photons. If Eve learns Alice's polarization (a measurement with one or more photons), she still sends a faked state and produces the expected interference at the beam splitter. The measure and reduce losses stage also gives the expected states and detector statistics. 

The statistics for $D_2$ inside Bob are the same as in the previous attack \eqref{eq:17}.

\subsection{Attack protocols against detectors with polarization discrimination in Bob's side and one of Alice's detector}
In some implementations, it might be the case that only one of Alice's detectors can distinguish the polarization of the incoming photons. As $D_2$ implicitly includes polarization discrimination, there is no difference in Bob's side for our attacks. However, Alice's detectors also see the light that Alice kept inside her delay line and polarization is indeed relevant.

The statistics at $D_2$ are still given by Eq. \eqref{eq:17}. On Alice's side it depends on which detector can distinguish different polarizations.

The proposed attacks follow the general protocol in the previous section with some differences. Now Eve can choose between completely blinding both of Alice's detectors or forcing a detection in the polarization sensitive detector and blind the other detector. Eve can send a blinding pulse at one of the polarizations and a smaller pulse at the orthogonal one (the polarization Eve wants to be detected). The polarizing beam splitter guarantees only the polarization sensitive detector clicks. Eve can use that to her advantage when she knows Alice's polarization choice.

\subsubsection{$D_2$ and $D_1$ can discriminate the polarization state}
This is the most natural scenario and has been implemented experimentally \cite{LJL12}. The photons that are used for the key are subject to an additional check to make sure their polarization is the expected one.
 
For her attack, Eve follows the protocol of the previous Section. All the steps are the same, but the detectors have a small change in their behaviour. When Alice's detectors are blinded in the polarization orthogonal to Bob's measuremente choice, the detector $D_0$, which cannot tell polarizations apart, is always blinded, irrespective of the choices of Alice and Bob.

If Bob measures the same polarization Alice chose, the detector statistics are:
\begin{equation}\label{eq:22}
\begin{aligned} 
{{P}^{3}_{D0}}&(n>0) = \frac{1}{2}(1-xe^{-{\eta_E}{T|\alpha|}^2})(1-e^{-{\eta_0}{\sigma^2}R^2|\alpha|^2})
\end{aligned}
\end{equation}
and
\begin{equation}\label{eq:21}
\begin{aligned} 
{P^{3}_{D1}}&(n>0) = \frac{1}{2}(1-e^{-{\eta_1}{\sigma^2}RT|\alpha|^2}).
\end{aligned}
\end{equation}

If Bob measures in a different polarization, $D_1$ never activates and $D_0$ is only blinded when Eve measures zero photons from Alice's state. The count probability for $D_0$ is:
\begin{equation}\label{eq:23}
\begin{aligned} 
{P'^{3}_{D0}}&(n>0) = \frac{1}{2}(1-xe^{-{\eta_E}{T|\alpha|}^2})(1-e^{-{\eta_0}{\sigma^2}{|\alpha |}^2}).
\end{aligned}
\end{equation}

\subsubsection{$D_2$ and $D_0$ can discriminate the polarization state}
While this a more artifical scenario, we include it for completeness. We follow the attack of Section \ref{poldisc}, which results in slighly different probabilities, with a small modification in the perfect knowledge scenario for the cases where there should be an interference in Alice's beam splitter. 

If Bob measures the same polarization Alice chose, the detector statistics are:
\begin{equation}
\begin{aligned} 
{{P}^{4}_{D0}}&(n>0) = \frac{1}{2}(1-e^{-{\eta_0}{\sigma^2}R^2|\alpha|^2})
\end{aligned}
\end{equation}
and
\begin{equation}
\begin{aligned} 
{P^{4}_{D1}}&(n>0) = \frac{1}{2}(1-xe^{-{\eta_E}{T|\alpha|}^2})(1-e^{-{\eta_1}{\sigma^2}RT|\alpha|^2}).
\end{aligned}
\end{equation}
Now, Eve can only blind $D_0$ for a particular polarization and $D_1$ is always blinded.

If Bob measures in a different polarization, only $D_0$ can activate. Eve can compensate for all the lost counts when she measures 0 photons and blinds Alice by selectively inducing a detection in $D_0$ with a probability $z_0$ in the perfect knowledge scenario. Eve induces the detection with a blinding pulse in the polarization that is not expected combined with a trigger pulse at the desired polarization. $D_1$ sees a blinding pulse and only the branch of $D_0$ in the right polarization is activated. The count statistics for $D_0$ are:
\begin{equation}
\begin{aligned} 
{P'^{4}_{D0}}&(n>0) = \frac{1}{2}(1-x)(1-e^{-{\eta_0}{\sigma^2}{|\alpha |}^2})\\
&\quad+\frac{1}{2}x(1-e^{-{\eta_E}{T|\alpha|}^2})z_0.
\end{aligned}
\end{equation}
Eve can adjust $z_0$ for every possible $x$. If the difference ${P'_{D0}}(n>0)-{P'^{4}_{D0}}(n>0)$ is greater than $\frac{1}{2} x(1-e^{-{\eta_E}{T|\alpha|}^2})$, Eve can set $z_0$ to 1 or adjust $z_0$ to match the expected probability otherwise. 
 
\section{Numerical analysis of the attack}
Eve must optimize her parameters so that the count statistics at each detector are as close as possible to those of an eavesdropper-free key exchange, as given in Eqs. (\ref{PD2}-\ref{PpD0}).

In this Section we show that, for different realistic parameters, the attack protocols we have presented can be successful. The efficiency of the attack is quantified using the ratios between the detector statistics under attack and the expected statistics: ${P^i_{D0}}/{P_{D0}}$, ${P^i_{D1}}/{P_{D1}}$, ${P^i_{D2}}/{P_{D2}}$ and ${P'^i_{D0}}/{P'_{D0}}$ (with $i=1,2,3,4$ indicating which attack protocol is used).

We performed a brute force search numerical optimization. The parameters $x$ and $z$ (no polarization discrimination) or $x$ (for polarization discriminating detectors) were explored to find the values that gave the smallest maximum deviation from 1 in the probability ratios, so that, even for the most sensitive detector statistics, the effect of the attack is small. The parameters $y$ and $z_0$ (when only $D_0$ distinguish polarization in Alice) can be computed exactly for each value of $x$ and $z$ (where applicable).

We start with a typical scenario with detector efficiency $\eta_0 = \eta_1 = \eta_2 = \eta_E = 0.1$ and a mean photon number from Alice $\left<n\right> = 0.1$ and consider a channel transmission of $\sigma = 0.1$, which corresponds to the transmission of a $50~\kilo\meter$ link of standard optical fiber with a $0.2~\deci\bel/\kilo\meter$ (ignoring all the other loss sources).

We assume Eve can improve the channel transmission to obtain a value $\sigma' = 1.2\sigma$, which only requires a $0.8~\deci\bel$ loss reduction which could come, for instance, from replacing $20~\kilo\meter$ of the channel fiber by a low loss fiber with $0.16~\deci\bel/\kilo\meter$ loss. The higher transmission is needed for the blind and reduce losses part of the attacks. 

Table \ref{TableR} shows the best ratios for an attack under different configurations of the counterfactual QKD system, including different beam splitting ratios inside Alice and the presence or absence of detectors that can distinguish the polarization of the photons. The tables include the value of the parameters $x$, $y$, $z$ and $z_0$ which minimize the maximum deviation from the ideal ratio of 1 for the worst of the four detector statistics.

\begin{table}[t!]
           \caption{Attack efficiencies for $\eta_0 = \eta_1 = \eta_2 = \eta_E = 0.1$, $\left<n\right> = 0.1$, $\sigma = 0.1$ and $\sigma' = 1.2\sigma$.\label{TableR}}
	\centering
\begin{tabular}{|C|C|C|C|}
\hline
\multicolumn{4}{|c|}{No polarization discrimination}     \\  \hline		
&  $R=0.5$ & $R=0.4$  & $R=0.1$  \\  \hline
${P^1_{D0}}/{P_{D0}}$ & 1.01383  & 1.02152 & 1.03706 \\ \hline
${P^1_{D1}}/{P_{D1}}$ & 1.01383 & 0.99747 & 0.96286\\ \hline
${P^1_{D2}}/{P_{D2}}$ & 0.99383 & 0.98426 & 0.96497\\ \hline
${P'^1_{D0}}/{P'_{D0}}$ & 0.98613 & 0.97848 & 0.96228\\ \hline
$x$ & 0.042  & 0.041 &  0.039 \\ \hline
$y$ &  1.0 & 1.0 & 1.0  \\ \hline
$z$ & 0.668  & 0.472  & 0.024  \\ \hline
\end{tabular}
\vspace{3ex}\\
\begin{tabular}{|C|C|C|C|}
\hline
\multicolumn{4}{|c|}{Polarization discrimination in $D_0$, $D_1$ and $D_2$}     \\  \hline		
&  $R=0.5$ & $R=0.4$  & $R=0.1$  \\  \hline
${P^2_{D0}}/{P_{D0}}$ &  1.0  & 1.0  &1.0 \\ \hline
${P^2_{D1}}/{P_{D1}}$ &  1.0 & 1.0 & 1.0 \\ \hline
${P^2_{D2}}/{P_{D2}}$ &  0.96570   &  0.96551 &  0.96497\\ \hline
${P'^2_{D0}}/{P'_{D0}}$ &  0.96119   & 0.96123 & 0.96135\\ \hline
$x$ & 0.039  & 0.039 &  0.039 \\ \hline
$y$ &  1.0 & 1.0 & 1.0  \\ \hline
\end{tabular}
\vspace{3ex}\\

\begin{tabular}{|C|C|C|C|}
\hline
\multicolumn{4}{|c|}{Polarization discrimination in $D_1$ and $D_2$ only}     \\  \hline		
&  $R=0.5$ & $R=0.4$  & $R=0.1$  \\  \hline
${P^3_{D0}}/{P_{D0}}$ &  0.96119 & 0.96123  & 0.96135 \\ \hline
${P^3_{D1}}/{P_{D1}}$ &  1.0 & 1.0 & 1.0 \\ \hline
${P^3_{D2}}/{P_{D2}}$ &  0.96570 & 0.96551 & 0.96497 \\ \hline
${P'^3_{D0}}/{P'_{D0}}$ &  0.96119 & 0.96123 & 0.96135 \\ \hline
$x$ & 0.039  & 0.039 &  0.039 \\ \hline
$y$ &  1.0 & 1.0 & 1.0  \\ \hline
\end{tabular}
\vspace{3ex}\\

\begin{tabular}{|C|C|C|C|}
\hline
\multicolumn{4}{|c|}{Polarization discrimination in $D_0$ and $D_2$ only}     \\  \hline		
&  $R=0.5$ & $R=0.4$  & $R=0.1$  \\  \hline
${P^4_{D0}}/{P_{D0}}$ &  1.0  & 1.0  &1.0 \\ \hline
${P^4_{D1}}/{P_{D1}}$ &  0.96119 & 0.96123 & 0.96135\\ \hline
${P^4_{D2}}/{P_{D2}}$ &  0.96570 & 0.96551 & 0.96497 \\ \hline
${P'^4_{D0}}/{P'_{D0}}$ &  1.0 & 1.0 & 1.0 \\ \hline
$x$ & 0.039  & 0.039 &  0.039 \\ \hline
$y$ &  1.0 & 1.0 & 1.0  \\ \hline
$z_0$ &  0.02005 &  0.01672 &  0.01116  \\ \hline
\end{tabular}
\end{table}

\begin{table}[t!]
           \caption{Attack efficiencies for $\eta_0 = \eta_1 = \eta_2  = 0.1$, $\left<n\right> = 0.1$ and $R=T=0.5$.\label{Table2}}
	\centering
\begin{tabular}{|C|C|C|C|}
\hline
\multicolumn{4}{|c|}{No polarization discrimination}     \\  \hline		
&  $\sigma = 0.6$, $\sigma' = 0.72$ and $\eta_E=0.9$ & $\sigma=\sigma' = 0.1$ and $\eta_E=0.1$ & $\sigma=\sigma' = 0.1$ and $\eta_E=0.9$  \\  \hline
${P^1_{D_0}}/{P_{D0}}$ &  1.00850  & 1.01680  & 1.00182 \\ \hline
${P^1_{D1}}/{P_{D1}}$ &  1.00850  & 1.01680  & 1.00182\\ \hline
${P^1_{D2}}/{P_{D2}}$ &  0.99433  & 0.98335 &  1.0\\ \hline
${P'^1_{D0}}/{P'_{D0}}$ &  0.99149   & 0.98315 & 0.99818 \\ \hline
$x$ & 0.028  & 0.051 &  0.006 \\ \hline
$y$ &  1.0 & 1.0 & 0.95236  \\ \hline
$z$ & 0.682  & 0.668  & 0.682  \\ \hline
\end{tabular}
\vspace{3ex}\\

\begin{tabular}{|C|C|C|C|}
\hline
\multicolumn{4}{|c|}{Polarization discrimination in $D_0$, $D_1$ and $D_2$}     \\  \hline	
&  $\sigma = 0.6$, $\sigma' = 0.72$ and $\eta_E=0.9$ & $\sigma=\sigma' = 0.1$ and $\eta_E=0.1$ & $\sigma=\sigma' = 0.1$ and $\eta_E=0.9$  \\  \hline	
${P^2_{D0}}/{P_{D0}}$ &  1.0  & 1.0  &1.0 \\ \hline
${P^2_{D1}}/{P_{D1}}$ &  1.0 & 1.0 & 1.0 \\ \hline
${P^2_{D2}}/{P_{D2}}$ &  0.98024   & 0.95492 &  1.0\\ \hline
${P'^2_{D0}}/{P'_{D_0}}$ &  0.97419   & 0.95224 & 0.99426 \\ \hline
$x$ & 0.027  & 0.048 &  0.006 \\ \hline
$y$ &  1.0 & 1.0 & 0.95236  \\ \hline
\end{tabular}
\vspace{3ex}\\

\begin{tabular}{|C|C|C|C|}
\hline
\multicolumn{4}{|c|}{Polarization discrimination in $D_1$ and $D_2$ only}     \\  \hline	
&  $\sigma = 0.6$, $\sigma' = 0.72$ and $\eta_E=0.9$ & $\sigma=\sigma' = 0.1$ and $\eta_E=0.1$ & $\sigma=\sigma' = 0.1$ and $\eta_E=0.9$  \\  \hline
${P^3_{D0}}/{P_{D0}}$ &  0.97419  & 0.95224  &0.99426 \\ \hline
${P^3_{D1}}/{P_{D1}}$ &  1.0 & 1.0 & 1.0 \\ \hline
${P^3_{D2}}/{P_{D2}}$ &  0.98024   & 0.95492 &  1.0\\ \hline
${P'^3_{D0}}/{P'_{D_0}}$ &  0.97419   & 0.95224 & 0.99426 \\ \hline
$x$ & 0.027  & 0.048 &  0.006 \\ \hline
$y$ &  1.0 & 1.0 & 0.95236  \\ \hline
\end{tabular}
\vspace{3ex}\\

\begin{tabular}{|C|C|C|C|}
\hline
\multicolumn{4}{|c|}{Polarization discrimination in $D_0$ and $D_2$ only}     \\  \hline	
&  $\sigma = 0.6$, $\sigma' = 0.72$ and $\eta_E=0.9$ & $\sigma=\sigma' = 0.1$ and $\eta_E=0.1$ & $\sigma=\sigma' = 0.1$ and $\eta_E=0.9$  \\  \hline
${P^4_{D0}}/{P_{D0}}$ &  1.0  & 1.0  &1.0 \\ \hline
${P^4_{D1}}/{P_{D1}}$ &  0.97419 & 0.95224 & 0.99426 \\ \hline
${P^4_{D2}}/{P_{D2}}$ &  0.98024   & 0.95492 &  1.0\\ \hline
${P'^4_{D0}}/{P'_{D_0}}$ & 1.0   & 1.0 & 1.0 \\ \hline
$x$ & 0.027  & 0.048 &  0.006 \\ \hline
$y$ &  1.0 & 1.0 & 0.95236  \\ \hline
$z_0$ &  0.08167 & 0.02005 &  0.00227  \\ \hline
\end{tabular}
\end{table} 

For the usual experimental choice of $R=T=0.5$ \cite{RWW11,BCD12,LJL12} the disturbance in the statistics is below 1.5\%.

The effect of the attacker can be attributed to channel variability or errors in the estimation of the channel loss. If we take Eqs. (\ref{PD2}-\ref{PpD0}) and compute the ratios between the expected statistics for the ideal channel and a channel with slightly higher or lower losses, the smallest variation in any of the four statistics is above the 1.5\% of the $R=0.5$ case for a fluctuation of $0.069~\deci\bel$ in the expected loss. Even for worst cases of the table, $R=0.5$ and polarization discrimination in $D_1$, the deviation is slightly below the 4\% and the effect of Eve can be explained as a fluctuation of $0.18~\deci\bel$ in the channel. This is a reasonable expectation for experimental implementations of counterfactual QKD where there might be small instabilities in the system equivalent to this loss variation during operation. The changes in the statistics due to the attacks are small enough to be reasonable attributed to changes in the fiber channel and connectors, small variations of the laser power or just to the unavoidable experimental error when establishing the expected channel loss. 
 
As expected, polarization discrimination makes the attack less succesful. In particular, detecting the polarization for the key generating detections (in $D_1$) can help to detect an eavesdropper.

Table \ref{Table2} gives the efficiency for three additional configurations that show the reach of the blinding attacks. The first columns correspond to a channel with a higher transmission ($\sigma=0.6$). Here, in order to minimize the effect of Eve, we assume she has access to detectors with an efficiency $\eta_E=0.9$. While these detectors are more expensive and not as common as the $\eta_E=0.1$ APDs we were assuming, they are a reasonable cost for a dedicated attacker and there exist different technologies which can be used \cite{Had09,EFM11}. As it happened in the previous examples, full polarization discrimination decreases the efficiency of the attack, but Eve can still remain below a 3\% disturbance in the detector statistics. 

Finally, we would like to remark that, for many parameters, blinding is enough to pull out a successful attack, even without reducing the channel loss, which makes the attack viable with minimal resources and channel intervention. The two last columns of Table \ref{Table2} show that, even if there is no loss reduction, Eve can keep her disturbance below a 5\% error. Using a better detector $\eta_E=0.9$ she can increase her information about Alice's state and reduce her effect on the detector statistics to less than a 1\%.

\section{Analysis, countermeasures and future work}
We have shown that, like other QKD protocols, counterfactual QKD is vulnerable to detector blinding attacks. Eve can learn Bob's internal state and make the whole key exchange insecure. Eve can hide her presence perfectly if she can reduce the channel losses by $3~\deci\bel$, giving a valid attack for both single photon and weak coherent state implementations. For coherent weak state implementations, it is possible to perform an attack protocol combining measurement, blinding and a moderate channel loss reduction, which is not strictly necessary in some scenarios. All the stages of the combined attack can be carried out with present technology.

While the fact that counterfactual QKD keeps part of the quantum state out of the channel gives a partial protection, it is not enough. Detector blinding must be explicitly addressed when designing a counterfactual QKD system. 

Blinding attacks have been extensively studied in other QKD protocols \cite{LJW11,GLL11,LLK14,QKM18,CHE19,SSL20} and there are many proposed countermeasures. A simple solution is adding a ``watchdog'' detector at the input of Alice and Bob using a higly transmissive beam splitter to sample a small amount of the input light and direct it to a light detector, which would discover the higher light levels used for blinding attacks, as well as some forms of Trojan horse attack \cite{GFK06,JAK14,JSK15,SMJ17}, which are also a concern for counterfactual QKD \cite{WZT12,YWM16}. In many proposals, this detector is classical, but it is important to make sure this is enough. For some detection schemes the blinding and trigger pulses can have as little as hundreds of photons \cite{LJW11}.

Other proposals like using intensity modulation, random variation of the detector efficiencies or advanced receiver configurations \cite{LWL15,LPW16,KDL18,KLD18} could also be adapted for their use in the counterfactual QKD protocol. In any case, these proposals should be evaluated with care to avoid slightly modified attacks that circumvent them, as it has happened in the past \cite{HSC16,LWW10c}. The best approach would be including a realistic model of the detector which includes all its faults in the security proofs.

Finally, it is worth noticing that the attack protocols we have proposed can be adapted with minimal changes for other hacking methods that control the detectors, like after-gate attacks \cite{WLW11}. As far as the attacks are concerned, the detector behaviour is equivalent and a practical attack protocol would only require a small change in the optical elements used or the timing characteristics.

\section*{Acknowledgments}
This work has been funded by Junta de Castilla y Le\'on (project VA296P18).

\newcommand{\noopsort}[1]{} \newcommand{\printfirst}[2]{#1}
  \newcommand{\singleletter}[1]{#1} \newcommand{\switchargs}[2]{#2#1}

\end{document}